\documentclass[12pt]{article}
\usepackage{jheppub}

\pdfoutput=1

\usepackage{amsmath,bbm,array,amsfonts,graphicx,wrapfig,lscape,float,mathtools,multirow,longtable}
\usepackage[dvipsnames]{xcolor}
\usepackage{array}
\usepackage{mdframed}

\newcommand{\be}{\begin{equation}}
\newcommand{\ee}{\end{equation}}
\newcommand{\beq}{\begin{equation}}
\newcommand{\beql}[1]{\begin{equation}\label{#1}}
\newcommand{\eeq}{\end{equation}}
\newcommand{\ba}{\begin{array}}
\newcommand{\ea}{\end{array}}
\newcommand{\bea}{\begin{eqnarray}}
\newcommand{\beal}[1]{\begin{eqnarray}\label{#1}}
\newcommand{\eea}{\end{eqnarray}}
\newcommand{\ben}{\begin{enumerate}}
\newcommand{\een}{\end{enumerate}}
\newcommand{\bean}{\begin{eqnarray*}}
\newcommand{\eean}{\end{eqnarray*}}
\newcommand{\eref}[1]{(\ref{#1})}
\newcommand{\sref}[1]{\S\ref{#1}}

\newcommand{\fref}[1]{Figure \ref{#1}}
\newcommand{\btab}[1]{\begin{tabular}{#1}}
\newcommand{\etab}{\end{tabular}}

\newcommand{\comment}[1]{}

\newcommand{\qed}{\nobreak \ifvmode \relax \else
      \ifdim\lastskip<1.5em \hskip-\lastskip
      \hskip1.5em plus0em minus0.5em \fi \nobreak
      \vrule height0.75em width0.5em depth0.25em\fi}

\definecolor{darkspringgreen}{rgb}{0.09, 0.45, 0.27}
\definecolor{forestgreen}{rgb}{0.13, 0.55, 0.13}

\usepackage{array}
\usepackage{comment}
\usepackage{physics}
\usepackage{subcaption}
\usepackage{tikz,tikz-3dplot}
\usepackage[colorlinks=true]{hyperref}
\newcolumntype{C}[1]{>{\centering\let\newline\\\arraybackslash\hspace{0pt}}m{#1}}

\definecolor{yellow2}{rgb}{0.98, 0.80, 0.20}

\title{A Hilbert Series for Generalized Toric Polygons}

\author[a,b]{Ignacio Carre\~no Bolla,}

\author[c,d,e]{Sebasti\'an Franco,}

\author[a,b]{Diego Rodr\'iguez-G\'omez}

\affiliation[a]{Department of Physics, Universidad de Oviedo \\  
C/ Federico Garc\'ia Lorca  18, 33007  Oviedo, Spain}
\affiliation[b]{Instituto Universitario de Ciencias y Tecnolog\'ias Espaciales de Asturias (ICTEA) \\
 C/~de la Independencia 13, 33004 Oviedo, Spain.}
\affiliation[c]{Physics Department, The City College of the CUNY\\
	160 Convent Avenue, New York, NY 10031, USA}
\affiliation[d]{Physics Program and \textsuperscript{$e$}Initiative for the Theoretical Sciences\\
	The Graduate School and University Center, The City University of New York\\
	365 Fifth Avenue, New York NY 10016, USA}

\emailAdd{ignaciocarbolla@gmail.com}	
\emailAdd{sfranco@ccny.cuny.edu}
\emailAdd{d.rodriguez.gomez@uniovi.es}

\abstract{We study the Hilbert series for $5d$ Superconformal Field Theories (SCFTs) engineered by Generalized Toric Polygons (GTPs), which extend the geometric realization of these theories from toric Calabi-Yau 3-folds to theories associated to general webs of 5- and 7-branes. Smoothed T-cones provide fundamental building blocks of GTP tessellations, generalizing the role of minimal triangles in toric diagrams. Building on this construction, we propose an extension of the Martelli-Sparks-Yau algorithm for computing Hilbert series of toric Calabi-Yau 3-folds that computes the Ehrhart series directly from GTP tessellations. The Ehrhart series is an invariant under Hanany-Witten transitions, which translate geometrically into polytope mutations.}

\begin{document}

\maketitle

\section{Introduction}

String and M-Theory provide powerful tools for studying 5-dimensional Superconformal Field Theories (SCFTs). Two of the most prominent constructions through which $5d$ SCFTs arise are M-theory on Calabi–Yau 3-fold (CY$_3$) singularities \cite{Intriligator:1997pq} and the low-energy limit of webs of intersecting $(p,q)$ 5-branes ending on 7-branes in Type IIB string theory \cite{Aharony:1997bh}. In the case of toric CY$_3$'s, there is a beautiful connection between these two scenarios: the web of 5-branes corresponds to the toric spine of the toric CY$_3$ \cite{Leung:1997tw}. 

However, not all $5d$ SCFTs engineered via $(p,q)$-webs can be realized through M-theory on a toric CY$_3$. The subclass of webs for which such a geometric engineering is possible consists of those in which every external 5-brane leg terminates on a distinct 7-brane.\footnote{A single $(p,q)$ fivebrane carries charges $(p,q)$ with ${\rm gcd}(p,q) = 1$.} In these cases, the $(p,q)$-web can be derived from the dual graph of a triangulated toric diagram. Specifically, the external edges of the toric diagram correspond to 5-branes ending on individual 7-branes. However, generic $(p,q)$-webs may feature multiple parallel 5-branes with identical $(p,q)$ charges ending on the same 7-brane. To efficiently describe such configurations, the concept of {\it generalized toric polygons} (GTPs) was introduced in \cite{Benini:2009gi}. GTPs generalize toric diagrams—which appear as special cases—and are typically represented using two types of dots, often depicted in black and white. A white node separating two parallel external edges indicates that the corresponding 5-brane legs are attached to the same 7-brane.
Only recently has significant progress been made in connecting GTPs with geometric engineering \cite{Bourget:2023wlb, Arias-Tamargo:2024fjt, CarrenoBolla:2024fxy}.

Broadly speaking, $(p,q)$-webs admit various types of deformations, each with a distinct interpretation in the $5d$ theory. For example, separating the web into sub-webs that slide along the 7-branes corresponds to motion along the Higgs branch. In addition, deformations of the web within its plane can be categorized into two types: those that preserve the positions of the external 7-branes, and those that modify them. Deformations that change the positions of the 7-branes correspond to relevant deformations. Such deformations trigger RG flows to an IR effective theory, which sometimes corresponds to a standard supersymmetric gauge theory, which in $5d$ is IR free. The former correspond to motion along the Coulomb branch of the 5d SCFT. For GTPs that are standard toric diagrams, these web deformations translate into ordinary triangulations. In the generic case, however, the opening of a $(p,q)$-web is more subtle, as it has to comply with the $s$-rule imposed by the external 7-branes \cite{Hanany:1996ie}. As discussed in \cite{Bergman:2020myx}, this can be taken into account by deforming the web into openings with triple intersections such that each of them has self-intersection -2. This ensures that the web is fully resolved into vertices that cannot be further decomposed, thereby indicating that no additional hidden degrees of freedom remain. Remarkably, these ``atomic” triple intersections have a beautiful geometric counterpart: the so-called {\it smoothed T-cones}, introduced in the mathematical literature beginning with \cite{Shepherd1988} and recently discussed in the context of $5d$ SCFTs in \cite{CarrenoBolla:2024fxy}. The analog of a triangulation for a toric diagram, in the case of generic GTPs, is a tessellation using smoothed T-cones \cite{CarrenoBolla:2024fxy}.\footnote{To be fully precise, the tessellations of GTPs generically require additional pieces that go beyond smooth T-cones, dubbed {\it locked superpositions} \cite{CarrenoBolla:2024fxy}. In this paper we will restrict to tessellations that do not contain them.} Indeed, the standard triangles used in toric diagrams are special cases of T-cones, and thus the conventional triangulation associated with a resolution of a toric CY$_3$ can be viewed as a particular instance of a GTP tessellation. When the $5d$ SCFT has a deformation to a gauge theory, relevant deformations can be regarded as supersymmetric VEVs for scalars in background vector multiplets coupled to the global symmetries of the gauge theory. For this reason, the combination of the Coulomb branch and relevant deformations is often referred to as the {\it extended Coulomb branch}.

An intrinsic feature of the engineering of $5d$ SCFTs via $(p,q)$-webs is that the same SCFT can often be realized through vastly different web configurations, related by Hanany-Witten (HW) transitions \cite{Hanany:1996ie}. More precisely, the 7-branes on which the web’s external legs terminate can move along those legs until they cross the web. As this happens, the branch cut associated with the moving 7-brane sweeps across part of the web, typically modifying the $(p,q)$ charges of some 7-branes and possibly leading to the creation or annihilation of branes. 

Understanding how to characterize the space of geometries associated to equivalent webs connected by HW transitions is essential for developing a complete picture of the geometric engineering of $5d$ SCFTs. In this context, smoothed T-cones play a central role: every collection of 5-brane legs terminating on a single 7-brane corresponds to a smoothed T-cone.\footnote{As previously mentioned, throughout this paper, we will focus on configurations without locked superpositions.} The geometric counterpart of a HW transition—known as a {\it polytope mutation} \cite{GalkinUsnich}—is realized as a flip of a smoothed T-cone about its apex, moving it from one side of the GTP to the other \cite{Arias-Tamargo:2024fjt}. In this process, and more broadly in the definition of GTPs, the choice of origin is a key consideration. 

As reviewed in \cite{Arias-Tamargo:2024fjt}, certain quantities remain invariant under mutations/HW transitions; among them the Hilbert series (HS).\footnote{It is important to stress that the specific form of Hilbert series depend on the grading. In Section \sref{section_HS_Toric_CY3} we will discuss alternative gradings. We will be interested in the special Hilbert series that is invariant under mutations.} Such invariants can, in turn, be exploited for the classification of mutation-equivalent geometries. Interestingly, the results of \cite{Arias-Tamargo:2024fjt, CarrenoBolla:2024fxy} indicate that this specific Hilbert series invariant can be extracted from the BPS quiver associated with the $5d$ SCFT.

In their seminal work \cite{Martelli:2006yb}, Martelli, Sparks and Yau (MSY) introduced a method for computing a fully refined HS for a toric CY$_3$ from a triangulation of its toric diagram. Since, more generally, GTPs are tessellated by smoothed T-cones, it is natural to ask whether the MSY prescription can be generalized to compute their HS by combining the contributions from individual smoothed T-cones. Moreover, it is reasonable to expect this problem to be connected to the observation that certain version of the HS is invariant under mutation/Hanany-Witten transitions (see \cite{Arias-Tamargo:2024fjt}). In this paper, we set out to explore these questions.

The organization of the paper is as follows. In Section \sref{section_HS_Toric_CY3}, we review the computation of the HS for toric CY 3-folds from geometric data and discuss various gradings of it. In Section \sref{section_HS_for_GTPs} we propose a HS for T-cones and for a class of GTPs. In Section \sref{section_HS_for_GTPs}, we propose a generalization of MSY prescription to compute the Ehrhart series of general GTPs tessellated by smoothed T-cones. We also discuss the invariance of the resulting HS under polytope mutation. We present our conclusions in \sref{section_conclusions}.

\section{The Hilbert Series for Toric CY$_3$'s}

\label{section_HS_Toric_CY3}

In \cite{Martelli:2006yb}, MSY introduced an elegant procedure for computing the HS of any toric CY$_3$, refined by three fugacities $(x,y,z)$ associated to its $U(1)^3$ global symmetry. The algorithm begins with a triangulation $\mathcal{T}$ of the toric diagram $P$ into elementary triangles of area $1/2$, such that $P = \bigcup_{a \in \mathcal{T}} T_a$. Each such triangle is defined by its vertices $V_1, V_2, V_3 \in \mathbb{Z}^3$, which, due to the CY condition, can be taken to have the form $V_i=(v_x^i,v_y^i,1)$. To every external edge of the triangle, we associate the outward pointing normal vector $(p_i,q_i)$ as well as the integer $h_i=\epsilon_{ijk}v_x^j v_y^k$.\footnote{To keep the notation concise, we use the indices $i,j,k=1,2,3$ to label both the vertices and the normal vectors of each triangle. We hope the distinction is clear from the context. Moreover, each vertex, vector, and their components should in principle carry a label $a$ indicating that they belong to a triangle $T_a$ in the triangulation. We also omit these labels to avoid cluttering the notation.} Note that $h_i$ can be interpreted as the signed area of a triangle $O V_j V_k$ on the $\mathbb{Z}^2$ plane $v_z=1$ with $O=(0,0,1)$. Since the edge $V_j V_k$ has lattice width $w = 1$ by construction, $h_i$ corresponds to the lattice height of the triangle, measured perpendicularly to the edge $V_j V_k$. To every triangle we then associate the contribution
\begin{equation}
        \mathbb{HS}_a=\prod_{i=1}^3 \frac{1}{(1- x^{p_i} y^{q_i} z^{h_i})}\,,  \label{HStriang}
\end{equation}
so that the HS of $P$ is
\begin{equation}
\label{HStriangulation}
\mathbb{HS}(x,y,z)=\sum_{a \in \mathcal{T}} \mathbb{HS}_a\,.
\end{equation}

For later purposes, it is useful to observe that in the contribution of each individual triangle in \eqref{HStriang}, each monomial of the form $(1 - x^{p_i} y^{q_i} z^{h_i})$ corresponds to the edge of the triangle that is dual to the $(p_i, q_i)$ leg.

\subsection{Unrefining the HS}

The HS we constructed above is a function of three fugacities $(x,y,z)$. For some applications, it is useful to consider certain unrefinements of it (see e.g. \cite{Bao:2024nyu} for a recent discussion).

\subsubsection{Superconformal R-Charge}

A natural version of the HS for a CY 3-fold is the one refined solely by the superconformal $R$-charge of the associated $4d$ $\mathcal{N}=1$ SCFT living on D3-branes probing the singularity. The superconformal $R$-charges can be determined using $a$-maximization in the gauge theory \cite{Intriligator:2003jj}. The $4d$ SCFT on the worldvolume of the D3-branes is holographically dual to the Type IIB string theory background $AdS_5\times X^5$, where $X^5$ is the $5d$ Sasaki-Einstein base of the CY$_3$ cone. This unrefinement of the HS is the generating function counting the mesonic gauge invariant operators of the theory graded by their superconformal $R$-charge \cite{Martelli:2005tp, Benvenuti:2006qr}.

This HS can also be computed geometrically. To do so, consider the function
\begin{equation}
\mathcal{V}=\frac{8\pi^3}{27}\mathbb{HS}[e^{-sb_1},\,e^{-sb_2},\,e^{-sb_3}]\,.
\end{equation}
This function admits a Laurent expansion around $s=0$ of the form
\begin{equation}
\mathcal{V}\sim \frac{C_3[b_{\alpha}]}{s^3}+\frac{C_2[b_{\alpha}]}{s^2}+\cdots\,,
\label{V_expansion_C3_C2}
\end{equation}
so that extremizing $C_3$ such that $C_3=C_2$ gives the $4d$ R-symmetry HS and is such that ${\rm Vol}[\mathcal{B}]=\lim_{s\rightarrow 0}s^3\mathcal{V}$ \cite{Eager:2010dk}. Since ${\rm Vol}\sim a^{-1}$, with $a$ the central charge of the $4d$ SCFT, this volume minimization procedure corresponds to $a$-maximization in the SCFT \cite{Martelli:2005tp}.

Let us illustrate this approach with the simplest example of $\mathbb{C}^3$. In this case, $P$ consists of a single triangle, whose vertices can be taken to be
\begin{equation}
V_1=(0,0,1)\,,\qquad V_2=(1,-1,1)\,,\qquad V_3=(0,-1,1)\,.
\end{equation}  
A straightforward application of the procedure above gives
\begin{equation}
\label{HStriangle}
\mathbb{HS}=\frac{1}{(1-x)\,(1-y)\,(1-\frac{1}{xy})}\,.
\end{equation}  
One can then see that
\begin{equation}
\frac{27}{8\pi^3}\mathcal{V}=-\frac{1}{b_1\,(b_1+b_2)\,(b_2+b_3)}\frac{1}{s^3}-\frac{b_3}{2b_2(b_1+b_2)\,(b_2+b_3)}\frac{1}{s^2}+\cdots\,.
\end{equation}

Imposing $C_3=C_2$ in the expansion \eqref{V_expansion_C3_C2} yields $b_3=2$, and extremizing with respect to $b_x$ and $b_y$ gives $(b_x,b_y)=(\frac{2}{3},-\frac{4}{3})$. This implies that ${\rm Vol}(\mathbb{S}^5)=\pi^3$, recovering the correct volume for the base of the cone, which in this case is simply $\mathbb{S}^5$.

\subsubsection{The Ehrhart Series}

\label{section_Ehrhart_Series}

Another possible unrefinement of the HS leads to the so-called Ehrhart series associated to the toric diagram \cite{mutations,Arias-Tamargo:2024fjt}. This choice HS is more relevant for our purposes, since it is invariant under the geometric counterpart of HW transitions.  

\begin{figure}[ht!]
	\centering
	\includegraphics[width=2.7cm]{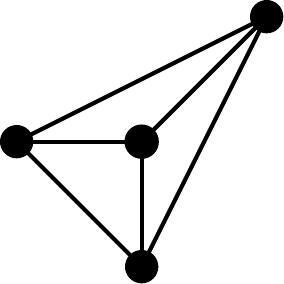}
\caption{Triangulated toric diagram for the complex cone over $dP_0$.}
	\label{Toric_dP0}
\end{figure}

For a toric diagram $P$ with internal points, the Ehrhart series is defined as
\begin{equation}
    {\rm Ehr}(P)=1+\sum_{n \geq 1}|n P^{\circ} \cap \mathbb{Z}^2| t^n\, ,
\end{equation}
where $P^{\circ}$ is the dual to $P$, that is
\begin{equation}
    P^{\circ}=\{u \in \mathbb{Q}^2,\, \forall v \in P, \, \langle u,v \rangle \geq 1\}\; .
\end{equation}
$P^{\circ}$ is a compact rational polygon if and only if the origin lies in the interior of $P$, so we will always choose the origin accordingly. For instance, consider the toric diagram with vertices
\begin{equation}
V_1=(-1,0)\,,\qquad V_2=(0,-1)\,,\qquad V_3=(1,1)\,,
\end{equation}
which corresponds to the complex cone over $dP_0$. This toric diagram is shown in \fref{Toric_dP0}.

The Ehrhart series is
\begin{equation}
\label{EhrdP0}
{\rm Ehr}=\frac{1+7t+t^2}{(1-t)^3}\,.
\end{equation}

Interestingly, the Ehrhart series can be obtained as an unrefinement of the HS computed as \eqref{HStriangulation}, which yields
\begin{equation}
\mathbb{HS}=\frac{1}{(1-y)\,(1-\frac{y}{x})\,(1-\frac{xz}{y^2})}+\frac{1}{(1-x)\,(1-\frac{x}{y})\,(1-\frac{yz}{x^2})}+\frac{1}{(1-\frac{1}{x})\,(1-\frac{1}{y})\,(1-xyz)}\,.
\end{equation}
Upon setting $x=y=1$, one obtains \eqref{EhrdP0}. One of the motivations of this paper is to provide a rationale for this unrefinement, which we will dub the {\it Ehrhart refinement}. As mentioned earlier, the Ehrhart refinement of the HS has the special property of being invariant under polytope mutation or, equivalently, HW transitions:

\medskip

\begin{mdframed}[linewidth=1pt, linecolor=black, backgroundcolor=gray!10, leftmargin=0pt, rightmargin=0pt, innertopmargin=10pt, innerbottommargin=10pt]

\noindent If $P$ and $P'$ are related by mutation (see \textit{e.g.} \cite{Arias-Tamargo:2024fjt} for a definition of mutation in the mathematical context)/HW in their associated $(p,q)$-webs, then the Ehrhart refinements of the HS of both $P$ and $P'$ coincide.
\end{mdframed}

\medskip

It is important to stress that the Ehrhart series depends on the origin, which is also crucial to identify the class of polytopes leading to equivalent Ehrhart series, as mutation can be regarded as flipping a T-cone (see below for a lightning review of $T$ cones and  \cite{Arias-Tamargo:2024fjt} for more details on mutation). More fundamentally, the complete specification of a $5d$ SCFT includes not just the polytope, but also a choice of origin—that is, the coordinates of its points. This combined data is what translates into a 5-brane web suspended from 7-branes \cite{Arias-Tamargo:2024fjt,CarrenoBolla:2024fxy}

\section{The HS for GTPs}

\label{section_HS_for_GTPs}

As reviewed in Section \sref{section_HS_Toric_CY3}, the HS for a toric CY 3-fold can be constructed from the elementary contributions of minimal area triangles in a triangulation of the underlying toric diagram \cite{Martelli:2006yb}. These individual contributions are then combined as in \eqref{HStriang}. The rationale is that, upon resolution, the space can be viewed as a union of $\mathbb{C}^3$ patches, each associated with a triangle in the triangulation. The formula \eqref{HStriang} assembles the contributions from each $\mathbb{C}^3$. It is natural to ask whether this procedure admits a generalization for arbitrary GTPs.

\subsection{T-Cones and the Coulomb Branch}

\begin{figure}[h!]
\centering
\includegraphics[scale=.4]{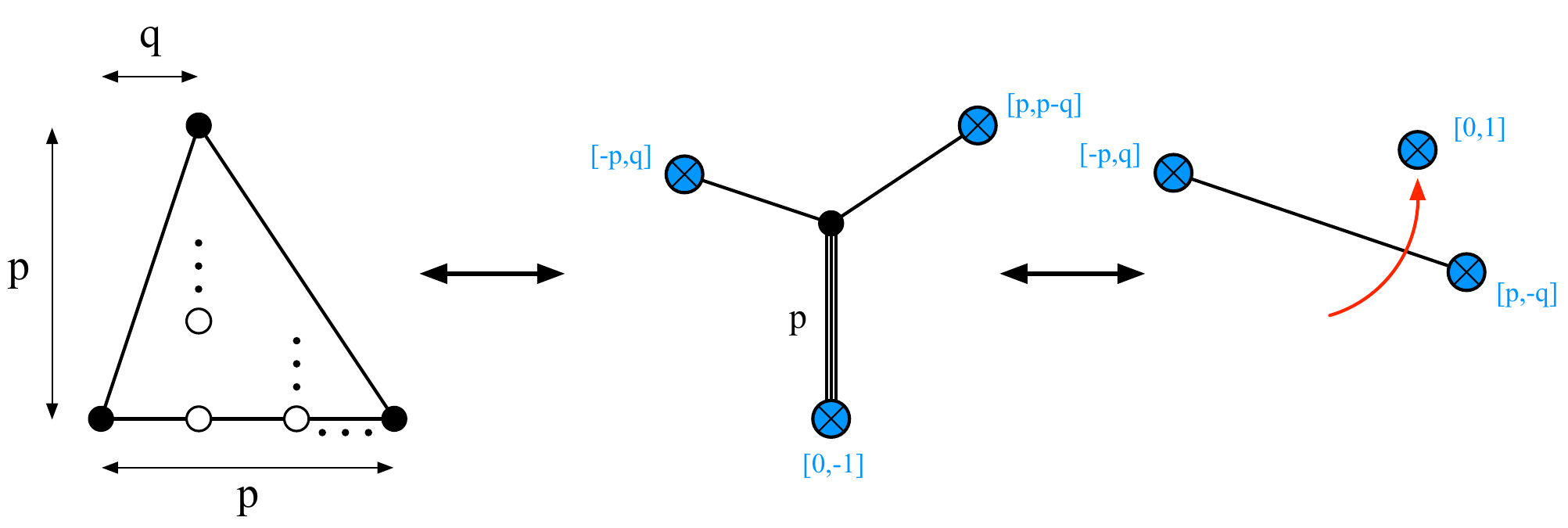}
\caption{GTP for a T-cone, its associated brane web and the detachment of one of the 7- branes upon a HW transition.}
\label{GTP-Web-T-cone}
\end{figure}

A generic $(p,q)$-web admits deformations corresponding to the extended Coulomb branch of the $5d$ SCFT. To fully display the entire extended Coulomb branch, one should maximally open the web until it consists of elementary triple intersections \cite{Bergman:2020myx}. These are triple intersections of 5-branes that cannot be further opened, so that they do not hide extra degrees of freedom. Such intersections admit a simple characterization as follows. Consider the triple intersection as an isolated configuration, represented by a Y-shaped $(p,q)$-web terminating on three 7-branes. Such an elementary triple intersection is characterized by the fact that, under a HW transition involving one of the legs, the configuration reduces to a single 5-brane stretched between two 7-branes, while the third 7-brane becomes detached. Conversely, such a configuration can be engineered by starting with a $(p,q)$ 5-brane stretched between two 7-branes and performing a crossing with a 7-brane. When considered in isolation, and up to an appropriate $SL(2,\mathbb{Z})$ transformation, these atomic triple intersections can always be brought to the canonical form $( p\,(0,-1) + (-p,q) + (p,p-q) )$, where we assume that $p$ and $q$ are coprime \cite{CarrenoBolla:2024fxy}. \fref{GTP-Web-T-cone} shows the corresponding GTP, the associated $(p,q)$-web, and the brane detachment via a HW transition that characterizes atomic junctions.

\begin{figure}[h!]
\centering
\includegraphics[scale=.4]{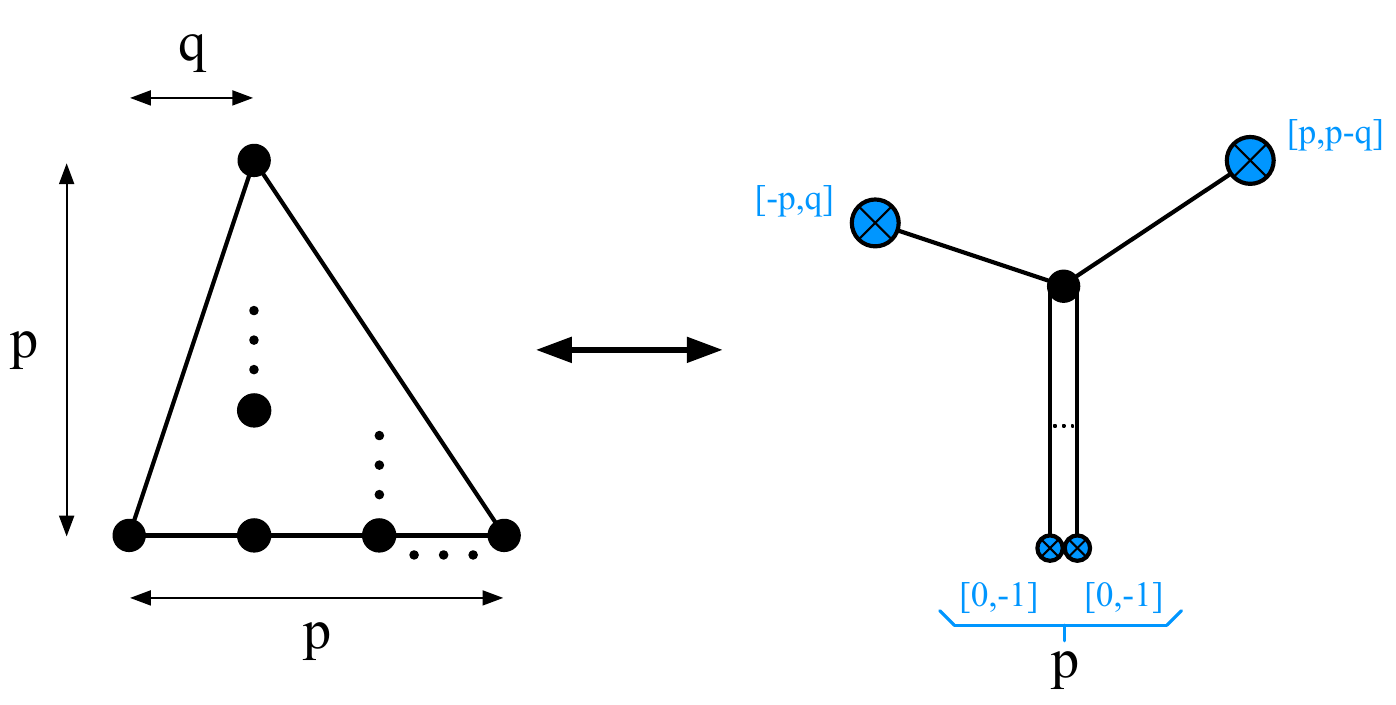}
\caption{A web in which each leg terminates on a distinct 7-brane is dual to a standard toric diagram, where all dots of the GTP become black.}
\label{Toric_Web_singular_T-cone}
\end{figure}

If, instead of all the vertical legs of the web terminating on a single 7-brane as in \fref{GTP-Web-T-cone}, each one ends on a different 7-brane, the resulting geometry no longer corresponds to a GTP with white dots, but rather to an ordinary toric diagram, as illustrated in \fref{Toric_Web_singular_T-cone}.

The toric diagram in \fref{Toric_Web_singular_T-cone} corresponds to a $\mathbb{C}^3/\mathbb{Z}_{p^2}$ orbifold with action
\begin{equation}
(z_1,\,z_2,\,z_3)\sim(\xi^p\,z_1,\,\xi^{p-q}\,z_2,\,\xi^{-2p+q}\,z_3)\,,\qquad \xi^{p^2}=1\,.
\end{equation}
This singularity is referred to as a {\it T-cone}. Geometrically, attaching all the vertical legs to a single 7-brane transforms the toric diagram in \fref{Toric_Web_singular_T-cone} into the GTP in \fref{GTP-Web-T-cone}, a process that corresponds to a smoothing of the singularity. The resulting geometry is known as a {\it smooth T-cone}.\footnote{For brevity, we will sometimes refer to both these objects simply as T-cones. Whether we mean the singular or smoothed version will be clear from context.} T-cones and their smoothings were first introduced in the mathematical literature in \cite{Shepherd1988}.

When $p$ and $q$ are coprime, as in our case, at most one side of the GTP or toric diagram contains more than one edge. Such configurations are known as {\it primitive T-cones}. Relaxing this condition leads to {\it non-primitive T-cones,} which can be regarded as Minkowski sums of primitive T-cones.

Returning to the analysis of the extended Coulomb branch using brane webs, the goal is to open them up into elementary triple intersections. Geometrically, this corresponds to tessellating the GTP with smoothed T-cones \cite{CarrenoBolla:2024fxy}.\footnote{As discussed in \cite{CarrenoBolla:2024fxy}, more general building blocks—corresponding to so-called locked superpositions—may be necessary to tessellate a generic GTP. In this paper, however, we restrict our attention to cases involving only smoothed T-cones.}

\subsection{HW Transitions, Polytope Mutations and T-cones}

HW transitions of a brane web translate, at the geometric level, into polytope mutations of the associated GTP. Polytope mutations admit both combinatorial and algebraic definitions \cite{mutations,Higashitani:2019vzu,Franco:2023flw,Arias-Tamargo:2024fjt}. The algebraic description is more refined, providing an explicit map between the coefficients of the Newton polynomials corresponding to the initial and final GTPs. Importantly, not all coefficients are in general independent: white dots indicate those that can be derived from others. The algebraic polytope mutation therefore allows us to describe HW transitions in the mirror geometry.

In a HW transition, a 7-brane moves to the opposite side of the web, thereby flipping its $(p,q)$ charge. As it crosses, other 7-branes may be transformed when they are swept through the branch cut of the flipped 7-brane. In addition, the HW brane-creation effect can lead to the appearance or disappearance of 5-branes. A 7-brane with $N$ 5-branes attached maps to a T-cone whose base length and lattice distance to the origin are both equal to $N$. In the dual language of polytope mutations, this T-cone is replaced by a new T-cone of opposite orientation and of size $N'$, which in general differs from $N$, thus capturing the effect of brane creation. The remainder of the polytope also undergoes a transformation to reflect the changes of the other 7-branes. We refer the reader to \cite{Franco:2023flw,Arias-Tamargo:2024fjt} for more detailed discussions of this process.

\subsection{The HS for Smoothed T-Cones}

As reviewed above, for generic GTPs—setting aside locked superpositions—the role of elementary triangles is taken over by T-cones. It is therefore natural to expect that T-cones also serve as the fundamental building blocks at the level of the HS. To explore this idea, we begin by examining T-cones in isolation to determine whether they exhibit any distinctive properties. Let us first consider the simplest genuine T-cone, whose vertices can be chosen to be
\begin{equation}
V_1=(0,0,1)\,,\qquad V_2=(-1,-2,1)\,,\qquad V_3=(1,-2,1)\, .
\label{vertices_example_T-cone_HS_1}
\end{equation}

\begin{figure}[ht!]
	\centering
	\includegraphics[width=6cm]{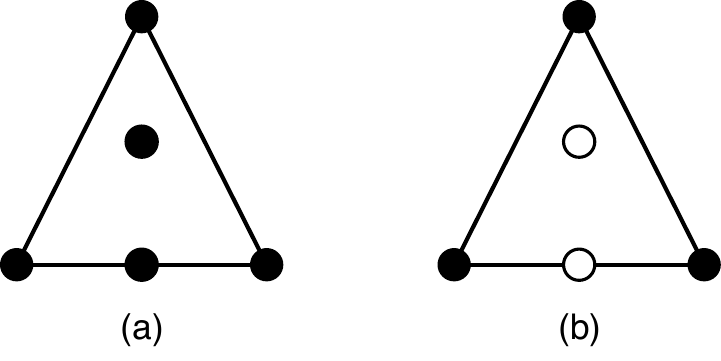}
\caption{a) Toric diagram and b) GTP for a primitive T-cone with vertices in \eqref{vertices_example_T-cone_HS_1} and its smoothing.}
	\label{Example_T-cone_HS_1}
\end{figure}

\fref{Example_T-cone_HS_1}.a shows the corresponding toric diagram, which can be triangulated using four elementary triangles. Using \eqref{HStriang}, we find
\begin{eqnarray}
\mathbb{HS}&=&\frac{1}{(1-\frac{1}{x})\,(1-\frac{x^2}{y})\,(1-\frac{yz}{x})}+\frac{1}{(1-x)\,(1-\frac{1}{x^2 y})\,(1-xyz)}\nonumber \\ &+& \frac{1}{(1-x)\,(1-\frac{1}{xyz})\,(1-yz^2)}+\frac{1}{(1-\frac{1}{x})\,(1-\frac{x}{yz})\,(1-yz^2)}\,.
\label{HS_example_1_1}
\end{eqnarray}
This toric diagram corresponds to a T-cone dual to a web with two $(0,-1)$ branes ending on two different $[0,-1]$ 7-branes. \fref{Example_T-cone_HS_1}.b shows the GTP for the corresponding smoothed T-cone, which is dual to a web where the two $(0,-1)$ branes end on the same $[0,-1]$ 7-brane. Since the vertical direction is associated to the $y$ fugacity, it is reasonable to relate the smoothing to the unrefinement $y=1$. Setting $y=1$ in \eqref{HS_example_1_1}, we obtain
\begin{equation}
\label{y=1}
\mathbb{HS}_{y=1}=\frac{1}{(1-x)\,(1-\frac{1}{x})\,(1-z)}\,.
\end{equation}
Interestingly, this contribution coincides with that of an elementary triangle, given by \eqref{HStriangle}, when evaluated in the unrefined case $y=1$. This observation suggests that \eqref{y=1} may represent the universal contribution of a smoothed T-cone.

Encouraged by this result, let us consider a bigger primitive T-cone, with vertices
\begin{equation}
V_1=(0,0,1)\,,\qquad V_2=(-1,-3,1),\,\qquad V_3=(2,-3,1)\,.
\label{vertices_example_T-cone_HS_2}
\end{equation}

\begin{figure}[ht!]
	\centering
	\includegraphics[width=8cm]{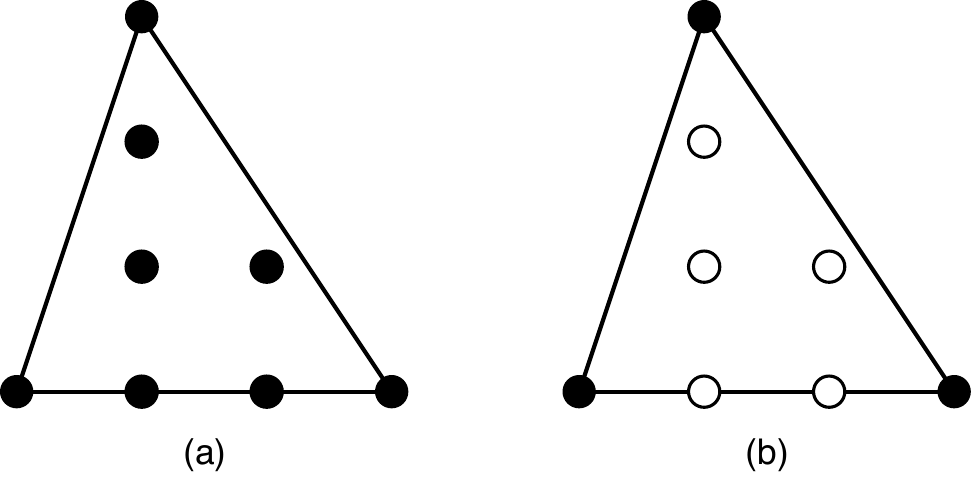}
\caption{a) Toric diagram and b) GTP for a primitive T-cone with vertices in \eqref{vertices_example_T-cone_HS_2} and its smoothing.}
	\label{Example_T-cone_HS_2}
\end{figure}

\fref{Example_T-cone_HS_2}.b shows the corresponding toric diagram. Although this toric diagram is triangulated into nine elementary triangles, it is straightforward to verify that, upon setting $y=1$, we once again recover \eqref{y=1}. This further suggests that \eqref{y=1} is the universal contribution to the HS associated to any smoothed primitive T-cone.

\begin{figure}[ht!]
	\centering
	\includegraphics[width=6cm]{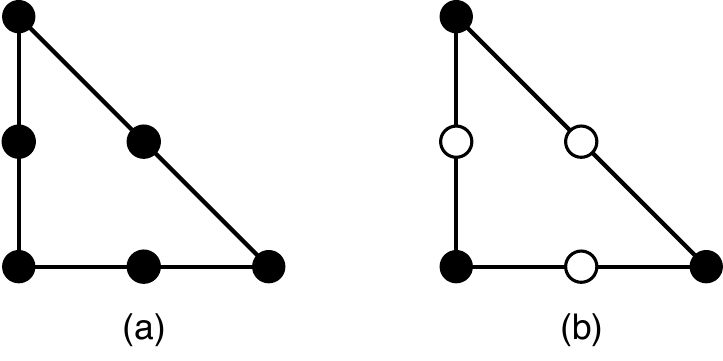}
\caption{a) Toric diagram and b) GTP for a non-primitive T-cone with vertices in \eqref{vertices_example_T-cone_HS_3} and its smoothing.}
	\label{Example_T-cone_HS_3}
\end{figure}

Let us now turn to the case of a non-primitive T-cone. As a representative example, consider a toric diagram whose vertices are given by
\begin{equation}
V_1=(0,0,1)\,,\qquad V_2=(0,-2,1)\,,\qquad V_3=(2,-2,1)\,,
\label{vertices_example_T-cone_HS_3}
\end{equation}
which we show in \fref{Example_T-cone_HS_3}.

In this case, \eqref{HStriang} gives
\begin{eqnarray}
\mathbb{HS}&=&\frac{1}{(1-x)\,(1-\frac{1}{xy})\,(1-yz)}+\frac{1}{(1-\frac{1}{xy})\,(1-\frac{x}{z})\,(1-yz^2)} \nonumber \\ &+& 
\frac{1}{(1-\frac{x}{yz^2})\,(1-\frac{z}{x})\,(1-yz^2)}+\frac{1}{(1-x)\,(1-\frac{1}{yz})\,(1-\frac{yz^2}{x})}\,.
\end{eqnarray}
Unrefining this expression by setting $y=1$, we find again \eqref{y=1}. 

\begin{figure}[ht!]
	\centering
	\includegraphics[width=2cm]{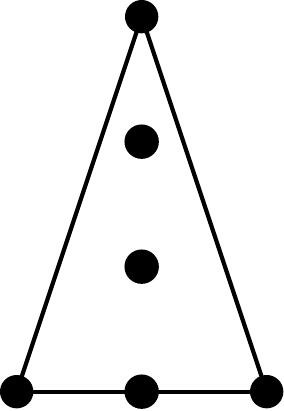}
\caption{Toric diagram with the vertices in \eqref{vertices_example_T-cone_HS_4}, which does not correspond to a T-cone.}
	\label{Example_T-cone_HS_4}
\end{figure}

In contrast, let us consider an example of a triangular toric diagram that does not correspond to a T-cone, as shown in \fref{Example_T-cone_HS_4}. This is evident from the fact that the length of the base differs from the height. The vertices of this diagram are located at
\begin{equation}
V_1=(0,0,1)\,,\qquad V_2=(-1,-3,1)\,,\qquad V_3=(1,-3,1)\,.
\label{vertices_example_T-cone_HS_4}
\end{equation}

Then, the MSY prescription \eqref{HStriang} gives
\begin{eqnarray}
\mathbb{HS}&=&\frac{1}{(1-\frac{1}{x})\,(1-\frac{x^3}{y})\,(1-\frac{yz}{x^2})}+\frac{1}{(1-x)\,(1-\frac{1}{x^3y})\,(1-x^2yz)} \nonumber \\
& +& \frac{1}{(1-\frac{1}{x})\,(1-\frac{x^2}{yz})\,(1-\frac{yz^2}{x})} + \frac{1}{(1-x)\,(1-\frac{1}{x^2yz})\,(1-xyz^2)} \nonumber \\ 
& + & \frac{1}{(1-x)\,(1-\frac{1}{xyz^2})\,(1-yz^3)}+\frac{1}{(1-\frac{1}{x})\,(1-\frac{x}{yz^2})\,(1-yz^3)}\,.
\end{eqnarray}
Trying the same unrefinement of previous examples, namely setting $y=1$, results in
\begin{equation}
\mathbb{HS}_{y=1}=\frac{x\,(1+2xz+z^2+x^2(1+z^2))}{(1-x)^2\,(1+x+x^2)\,(z^3-1)}\,,
\end{equation}
which is not of the form \eqref{y=1}, indicating that this simple form is only attainable for T-cones.

So far, we have aligned all primitive T-cones such that the side of the toric diagram containing more than one edge is parallel to the $x$-axis. In this configuration, the parallel legs of the dual web that eventually terminate on the same 7-brane upon smoothing are $(0,-1)$ 5-branes. Consequently, the unrefinement leading to the smoothed T-cones has simply been $y = 1$ in all these examples.

Using an appropriate $SL(2,\mathbb{Z})$ transformation, we can orient these edges so that they correspond to generic $(p,q)$ 5-brane legs. In this more general frame, the new variables are $(x', y', z') = (x, y^{-p/q}, z)$. One can verify that, in this generic case, the HS for smoothed T-cones is again given by expression in \eqref{y=1} and is attained by setting $y' = 1$.

Putting these observations together, we conjecture that the HS for a smoothed T-cone whose apex is at the origin is 

\begin{equation}
\label{sT-cone}
\mathbb{HS}_T=\frac{1}{(1-x)\,(1-\frac{1}{x})\,(1-z)}\,.
\end{equation}
Even though we do not offer a first principles proof of \eqref{sT-cone}, we have extensively checked its validity over many examples.

It is interesting to observe that the contribution in \eqref{sT-cone} can be obtained directly from the external vertices of the T-cone. Indeed, taking the vertices at

\begin{equation}
V_1=(0,0,1)\,,\qquad V_2=(-q,-p,1)\,,\qquad V_3=(p-q,-p,1)\,,
\end{equation}
eq. \eqref{HStriang} gives

\begin{equation}
\label{triangles}
\mathbb{HS}=\frac{1}{(1-x^{\frac{p}{{\rm gcd}(p,q)}})\,(1-\frac{1}{x^{\frac{p}{{\rm gcd}(p,q)}}})\,(1-z^p)}\,.
\end{equation}
Thus we can obtain the contribution in \eqref{sT-cone} directly from the external vertices of the T-cone upon re-scaling $(x,z)\rightarrow (x^{\frac{{\rm gcd}(p,q)}{p}},\,z^{\frac{1}{p}})$.

\subsection{The HS for GTPs}

We now turn to the problem of determining the HS for more general GTPs, in particular those characterized by multiple contributions arising from a tessellation. To make progress, we will concentrate on a special subclass of GTPs defined as follows:
\begin{itemize}
\item The GTP only contains smoothed T-cones (and no locked superpositions).
\item All smoothed T-cones share a common apex, which we place at the origin of the lattice, and all external sides with $p>1$ edges are aligned parallel to a given direction, which we take to be the $x$-axis.
\end{itemize}
We denote this class of GTPs by $\widehat{\rm GTP}$. Employing \eqref{sT-cone}, we can express the analog of \eqref{HStriang} for this class, which is given by:
\begin{itemize}
\item All minimal triangles contribute \eqref{HStriang} with $y=1$.
\item All smoothed T-cones contribute \eqref{sT-cone}.
\end{itemize}

\begin{figure}[ht!]
	\centering
	\includegraphics[width=2cm]{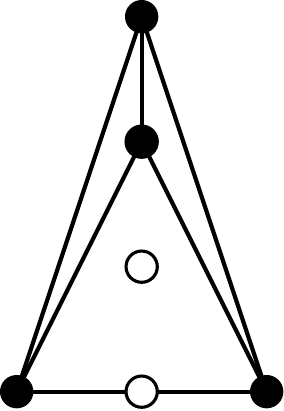}
\caption{GTP with the vertices in \eqref{vertices_example_T-cone_HS_5} and its triangulation.}
	\label{Example_T-cone_HS_5}
\end{figure}

For instance, using this recipe, let us consider a GTP closely related to the one in \fref{Example_T-cone_HS_4}, but in which every side of the polytope corresponds to a single D7-brane. For this to happen, the vertices should instead be
\begin{equation}
V_1=(0,1,1)\,,\qquad V_2=(-1,-2,1)\,,\qquad V_3=(1,-2,1)\,.
\label{vertices_example_T-cone_HS_5}
\end{equation}
\fref{Example_T-cone_HS_5} shows the corresponding GTP and its triangulation.

Using our prescription, the HS becomes
\begin{equation}
\label{HSY33}
\mathbb{HS}=\frac{1}{(1-x)\,(1-x^2)\,(1-\frac{z}{x^3})}+\frac{1}{(1-\frac{1}{x^2})\,(1-\frac{1}{x})\,(1-x^3z)}+\frac{1}{(1-x)\,(1-\frac{1}{x})\,(1-z)}\,.
\end{equation}

An interesting property of this result is that, if we unrefine it by setting $x=1$, it becomes
\begin{equation}
\label{HSY33}
\mathbb{HS}_{x=y=1}=\frac{1+7z+z^2}{(1-z)^3}\,.
\end{equation}
This is precisely the HS with the Ehrhart refinement for the cone over $dP_0$, which was given in \eref{EhrdP0}. This is not an accident, since, upon mutation, the GTP in \fref{Example_T-cone_HS_5} becomes the one for the cone over $dP_0$ (see \textit{c.f.} Figures 22.a and 22.b of \cite{Arias-Tamargo:2024fjt} for the transition between these polytopes, up to  $SL(2,\mathbb{Z})$ equivalence).

\section{A Choice of Origin, the Ehrhart Series and HW Transitions}

\label{section_proposal_HS}

The final observation of the previous section is very suggestive: unrefining by setting $x=y=1$ we have found the HS which coincides with the Ehrhart series. Let us examine this example in further detail. The HS for the cone over $dP_0$ was computed in Section \sref{section_Ehrhart_Series}. We reproduce it here for convenience
\begin{equation}
\label{dP0}
\mathbb{HS}=\frac{1}{(1-y)\,(1-\frac{y}{x})\,(1-\frac{xz}{y^2})}+\frac{1}{(1-x)\,(1-\frac{x}{y})\,(1-\frac{yz}{x^2})}+\frac{1}{(1-\frac{1}{x})\,(1-\frac{1}{y})\,(1-xyz)}\,.
\end{equation}
We observe that each individual term corresponding to a triangle in the triangulation diverges upon the unrefining $x=y=1$, yet the divergences cancel pairwise. The vanishing factors in the denominators of \eqref{dP0} are associated with the internal sides of the triangles, with each of them appearing exactly twice and with opposite orientations. As a result, the divergences cancel and the limit $x=y=1$ yields the finite result \eref{EhrdP0}. This suggests the following prescription for constructing the Ehrhart series:

\medskip

\begin{mdframed}[linewidth=1pt, linecolor=black, backgroundcolor=gray!10, leftmargin=0pt, rightmargin=0pt, innertopmargin=10pt, innerbottommargin=10pt]

\noindent For a given choice of an internal point as the origin, the Ehrhart refinement is obtained by applying the generalization of MSY formula \eqref{HStriangulation} to a triangulation of the GTP and unrefining the resulting HS by setting $x=y=1$.
\end{mdframed}

\medskip

\noindent Note that the GTP's are in general non-toric and consequently their HS is not expected to depend on 3 fugacities, in line with the unrefinement by setting $x=y=1$.

Let us test this proposal in various examples.

\begin{enumerate}
\item Complex cone over $\mathbb{F}_0$. The toric diagram is given in \fref{Toric_F0}. 

\begin{figure}[ht!]
	\centering
	\includegraphics[width=2.7cm]{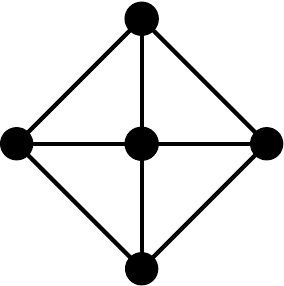}
\caption{Triangulated toric diagram for the complex cone over $\mathbb{F}_0$.}
	\label{Toric_F0}
\end{figure}

In this case there is a single internal point. Fixing it to the origin, the external vertices are
\begin{equation}
V_1=(-1,0,1)\,,\qquad V_2=(0,-1,1)\,,\qquad V_3=(1,0,1)\,,\qquad V_4=(0,1,1)\,.
\end{equation}
Computing the HS through \eqref{HStriangulation} and setting $x=y=1$, one finds

\begin{equation}
\label{F0}
\mathbb{HS}_{x=y=1}=\frac{1+6z+z^2}{(1-z)^3}\,,
\end{equation}
in agreement with the expectation from the Ehrhart series (see \textit{e.g.} \cite{Arias-Tamargo:2024fjt}).

\item Complex cone over $\mathbb{F}_2$. The toric diagram is given in \fref{Toric_F2}. 

\begin{figure}[ht!]
	\centering
	\includegraphics[width=2.7cm]{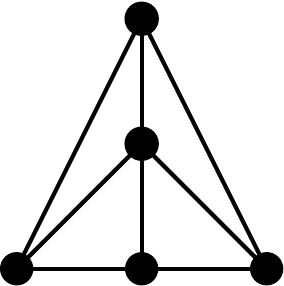}
\caption{Triangulated toric diagram for the complex cone over $\mathbb{F}_2$.}
	\label{Toric_F2}
\end{figure}

There is a single internal point. Fixing it to the origin, the external vertices are
\begin{equation}
V_1=(-1,-1,1)\,,\qquad V_2=(1,-1,1)\,,\qquad V_3=(1,0,1)\,,\qquad V_4=(0,1,1)\,.
\end{equation}
Computing the HS through \eqref{HStriangulation} and setting $x=y=1$ one finds 
\begin{equation}
\label{F2}
\mathbb{HS}_{x=y=1}=\frac{1+6z+z^2}{(1-z)^3}\,,
\end{equation}
which agrees with the Ehrhart series. Moreover, it coincides with \eqref{F0}, as expected, since the two cases are related by polytope mutation (see \textit{e.g.} \cite{Arias-Tamargo:2024fjt}).

\item Complex cone over $Y^{3,3}$. This is the example in \fref{Example_T-cone_HS_4}, which has two internal points. Choosing the upper one as the origin, we get the GTP in \fref{Example_T-cone_HS_5}. Applying our prescription leads to \eqref{EhrdP0} as discussed above. 

In turn, choosing the lower point as origin would correspond to vertices located at
\begin{equation}
V_1=(0,2,1)\,,\qquad V_2=(-1,-1,1)\,,\qquad V_3=(1,-1,1)\,.
\label{vertices_Y33_2}
\end{equation}
The corresponding GTP and its triangulation is shown in \fref{Toric_Y33_2}.

\begin{figure}[ht!]
	\centering
	\includegraphics[width=2cm]{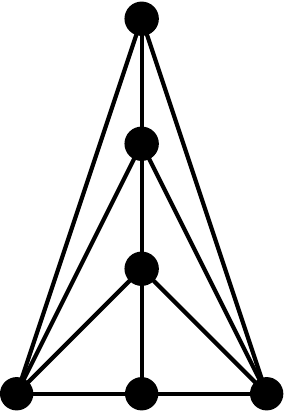}
\caption{GTP with the vertices in \eqref{vertices_Y33_2} and its triangulation.}	
\label{Toric_Y33_2}
\end{figure}

In this case upon setting, the prescription gives
\begin{equation}
\mathbb{HS}_{x=y=1}=\frac{1+3z+10z^2+3z^3+z^4}{(1-z)^3\,(1+z)^2}\,,
\end{equation}
which agrees with the Ehrhart series (\textit{cf.} equation (6.11) in \cite{Arias-Tamargo:2024fjt}).

\end{enumerate}

\section{Conclusions}

\label{section_conclusions}

This paper advances the geometric understanding of $5d$ SCFTs that can alternatively be engineered using general webs of 5- and 7-branes. Specifically, we introduced a generalization of the MSY prescription that enables the determination of the Ehrhart series directly from a tessellation of a GTP by smoothed T-cones. This provides strong support for the picture introduced in \cite{CarrenoBolla:2024fxy}. Quantitatively, at the level of the HS, the Ehrhart refinement can be understood as a union of smooth patches, each corresponding to a smoothed T-cone, in a way that is invariant under HW transitions/polytope mutations. Moreover, the contribution of every smoothed T-cone is identical, matches that of the familiar triangles for toric $CY_3$ in the MSY prescription and can be fully characterized in terms of the positions of the external vertices (\textit{cf.} \eqref{sT-cone} and \eqref{triangles}).

At this stage, our algorithm for obtaining the HS in the Ehrhart refinement should be regarded as a purely mathematical construction. Nevertheless, it is natural to conjecture that a deeper connection to the underlying BPS quiver may exist. Indeed, as demonstrated in several explicit examples in \cite{Arias-Tamargo:2024fjt}, with a suitable choice of grading the HS computed as the partition function of BPS operators in the BPS quiver coincides with the HS in the Ehrhart refinement. Uncovering the rationale for this grading and clarifying its relation to the algorithm proposed here is an interesting open problem that we plan to revisit.

Finally, while we have checked our proposal in a multitude of examples, it would be highly desirable to have an analytic proof. Moreover, looking ahead, we expect to extend this algorithm to general tessellations of GTPs, namely those including locked superpositions.

\section*{Acknowledgements}

S.F. was supported by the U.S. National Science Foundation grants PHY-2112729 and PHY-2412479. He would also like to thank the 2025 Simons Physics Summer Workshop and the Simons Center for Geometry and Physics for their hospitality during part of this work. I.C.B and D.R.G are supported in part by the Spanish national grant MCIU-22-PID2021-123021NB-I00. The work of I.C.B is also supported by the Severo Ochoa fellowship NAC-AT-PUB-ASV-2025 BP24-116.

\bibliographystyle{JHEP}
\bibliography{biblio.bib}

\end{document}